\documentclass{article}
\usepackage{dcase2024_techrep,amsmath,graphicx,url,times,booktabs, tabularx, bbding}

\title{Squeeze-and-Excite ResNet-Conformers for Sound Event Localization, Detection, and Distance Estimation for DCASE2024 Challenge}

\name{Jun Wei Yeow$^{1}\sthanks{This research is supported by the Ministry of Education, Singapore, under its Academic Research Fund Tier 2 (MOE-T2EP20221-0014)}$,
      Ee-Leng Tan$^{1}$,
      Jisheng Bai$^{2}$, 
      Santi Peksi$^{1}$,
      Woon-Seng Gan$^{1}$ 
      }
\address{$^1$ Smart Nation TRANS Lab, Nanyang Technological University, Singapore \\
junwei004@e.ntu.edu.sg, \{etanel, speksi, ewsgan\}@ntu.edu.sg\\          
        $^2$ School of Marine Science and Technology, Northwestern Polytechnical University, Xi'an, China\\ 
        baijs@mail.nwpu.edu.cn\\
 }

\begin{document}

\ninept
\maketitle

\begin{sloppy}

\begin{abstract}
This technical report details our systems submitted for Task 3 of the DCASE 2024 Challenge: Audio and Audiovisual Sound Event Localization and Detection (SELD) with Source Distance Estimation (SDE). We address only the audio-only SELD with SDE (SELDDE) task in this report. We propose to improve the existing ResNet-Conformer architectures with Squeeze-and-Excitation blocks in order to introduce additional forms of channel- and spatial-wise attention. In order to improve SELD performance, we also utilize the Spatial Cue-Augmented Log-Spectrogram (SALSA) features over the commonly used log-mel spectra features for polyphonic SELD. We complement the existing Sony-TAu Realistic Spatial Soundscapes 2023 (STARSS23) dataset with the audio channel swapping technique and synthesize additional data using the SpatialScaper generator. We also perform distance scaling in order to prevent large distance errors from contributing more towards the loss function. Finally, we evaluate our approach on the evaluation subset of the STARSS23 dataset. 
\end{abstract}


\section{Introduction}
\label{sec:intro}

The sound event localization and detection (SELD) task for the DCASE 2024 Challenge now introduces the additional task of sound distance estimation (SDE). Not only are systems required to perform sound event detection (SED) and its corresponding direction of arrival (DOA) estimation, systems now also need to estimate the distance between the recording device and the active sound source. The introduction of the SDE task into the challenge problem makes the task significantly more complex. Although there is a wealth of literature relating to SELD and SDE tasks separately, little work has been done on the joint SELD and SDE (SELDDE) task. In \cite{krause2024sound_dcasebaseline}, Krause et al. introduced the first form of investigation into this newly modeled SELDDE problem. The authors extend the existing multi activity-coupled Cartesian distance of arrival (multi-ACCDOA) vector proposed by Shimada et al. \cite{shimada_multiaccdoa} to include distance estimation, which they term the multi activity-coupled Cartesian distance and DOA (multi-ACCDDOA) method. The authors then modify the existing convolutional recurrent neural network (CRNN) SELDNet, which is used for many SELD tasks, to concurrently estimate the presence of active sound events and their corresponding spatial location. This modified CRNN is then used as the baseline system for DCASE 2024 Challenge Task 3.

CRNNs have also formed the basis of many state-of-the-art (SOTA) work in SELD. Most of the top-ranking submissions in the DCASE 2023 Challenge Task 3 utilize some form of CRNN-based architecture \cite{Kang_KT_task3_report_rank4_dcase2023, Du_NERCSLIP_task3_report, Bai_JLESS_task3a_report}, with the most common being the ResNet-Conformer system. We adopt the ResNet-Conformer architecture for our approach, using the multi-ACCDDOA format as its output. To further enhance SELD performance, we use the highly effective Spatial Cue-Augmented Log-Spectrogram (SALSA) features proposed by Tho et al. \cite{Nguyen2022_SALSA}. As the SALSA features utilize more frequency bins as it does not use the traditional Mel frequency bins, we include Squeeze-and-Excitation blocks \cite{hu2018squeeze, roy2018concurrent} to introduce forms of spatial attention to make full use of the rich amount of spatial information provided by the SALSA features. In addition, we use data augmentation techniques adapted from \cite{wang2023four} and synthesize additional data using the SpatialScaper generator \cite{Roman2024_SpatialScaper} to compliment the STARSS23 dataset \cite{politis_2023_starss23}. In general, we show that our approach achieves significant improvements over the baseline system without model ensembling.

\section{Proposed Method}
\label{sec:format}

\subsection{Feature Extraction}

The STARSS23 dataset provides audio data in both the first-order Ambisonics (FOA) and tetrahedral microphone (MIC) formats. We utilize only the FOA array signals in our proposed approach. We extract the SALSA features in a similar manner as detailed in \cite{Nguyen2022_SALSA}. We used a sampling rate of 24kHz, a window length of 512 samples, hop size of 300 samples, 512-point FFT and the Hann window. Similarly, we also linearly compress frequency bands above 9kHz to reduce the feature dimensions to reduce the training time. This results in an input shape of $(7 \times 400 \times 200)$ for number of input channels, time bins and frequency bins respectively.  

The SED, DOA, and SDE target labels were extracted and converted to the multi-ACCDDOA format. In \cite{krause2024sound_dcasebaseline}, the authors note that larger distances may contribute more significant errors. To mitigate this issue, we first standardize and then scale the distance distribution such that all distance values lie within $[-1, 1]$. Moreover, as all elements in the multi-ACCDDOA vector are now lie within the same range of $[-1, 1]$, we can also apply the \textit{tanh} operation to the model's output logits. This helps to introduce a further form of non-linearity to the system. The distance scaling method consists of two steps -- standardization followed by scaling. The standardization of the distance distribution is given by

\begin{equation}
    \label{eqn:dist_stand}
    d_{stand} = \frac{d -\Bar{d}}{\sigma_{d}},
\end{equation}

where \textit{d} represents the distance distribution in the data set in meters, with $\Bar{d}$ and $\sigma_d$ representing the mean and standard deviation of this distribution, respectively. This operation yields us the standardized distribution of distance values, $d_{stand}$, with approximately zero mean and unit standard deviation. The next step is to scale the distance distribution to $d_{scaled}$, such that $d_{scaled} \in [-1, 1]$. The following equation shows the scaling operation,

\begin{equation}
    \label{eqn:dist_scale}
    d_{scaled} = \frac{d_{stand}}{\text{max}(d_{stand})} \cdot
\end{equation}

To retrieve the original distance values in meters, we simply perform (\ref{eqn:dist_stand}) and (\ref{eqn:dist_scale}) in reverse order to yield the original $d$. We store $\Bar{d}, \sigma_d$ and $\text{max}(d_{stand})$ as variables to use to restore the distance values back from scaled units into meters during post-processing. However, we note that using this scaling method will limit our output distance values in meters to be within the minimum and maximum of the original \textit{d}. In this case, our output distance values are limited to $(0.04\text{m}, 7.64\text{m})$.  

\subsection{Dataset}
\label{sec:dataset}

The development set of the STARSS23 dataset consists of 168 recording clips recorded in 16 unique rooms. The training split of this development set only consists of 98 recordings captured across 12 rooms, totaling up to around 5.5 hours. In order to compliment the relatively small dataset provided, the DCASE Challenge organizers also provide an external synthesized dataset\footnote{https://zenodo.org/records/10932241} generated using real spatial room impulse responses (SRIRs) from the TAU-SRIR database \cite{politis_2022_6408611_tau_srir_db}. We refer to this dataset as the ``DCASE synthetic dataset". The DCASE synthetic dataset consists of 1200 one-minute long spatialized recordings, and we note that the dataset has its own imbalance in terms of class and distance distributions. As a result, we opt to generate our own set of audio recordings instead.

Due to the difficulty of manually annotating audio data, the STARSS23 dataset is limited by its size and distribution of classes. We address this problem by spatializing additional training data using the SpatialScaper generator \cite{Roman2024_SpatialScaper}. When synthesizing this additional dataset, we make special note of the SRIRs used to generate the audio data. Under the assumption that the rooms in the STARSS23 dataset are of similar size, we only select SRIRs that were recorded at similar distances to the sound source distances recorded in the development set of the STARSS23 dataset. We also focus on synthesizing sound event classes that are under-represented in the STARSS23 dataset, such as ``knock" or ``doorCupboard". We follow the synthesizing configurations of the provided additional dataset and limit the maximum polyphony to 3. We synthesized around 18 hours worth of additional data using SRIRs from 8 different rooms, which we refer to as the ``SpatialScaper dataset". During training, we opt to use the SpatialScaper dataset to compliment the STARSS23 training set as opposed to the DCASE synthetic dataset. We show the distribution of distances in the DCASE synthetic dataset and the SpatialScaper dataset compared to the STARSS23 training set in Figure \ref{fig:distance_dist}. 

\begin{figure}[t]
    \centering
    \includegraphics[width=0.49\columnwidth]{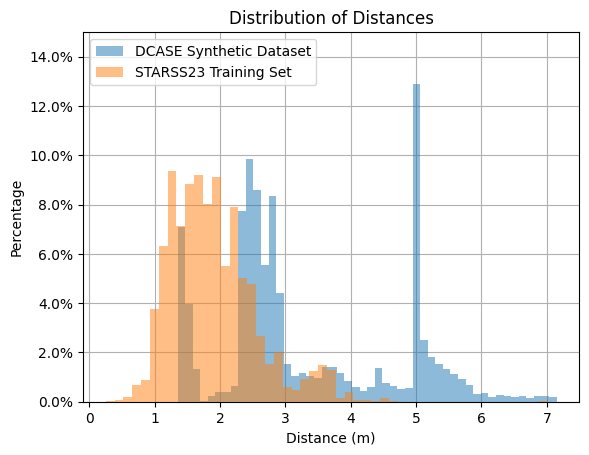}
    \includegraphics[width=0.49\columnwidth]{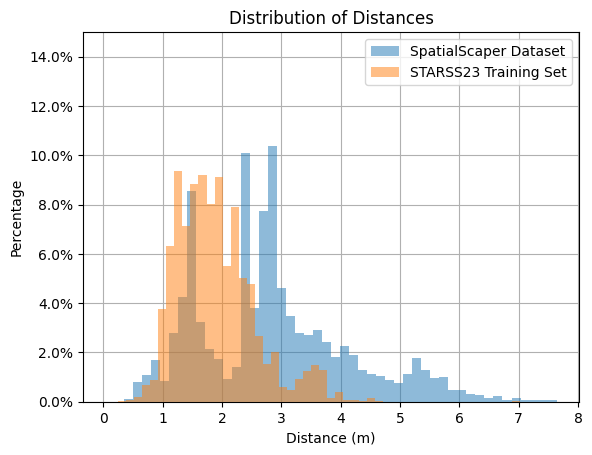}
    \caption{Distribution of distances within each dataset in comparison with STARSS23 training set}
    \label{fig:distance_dist}
\end{figure}

\subsection{Data Augmentation}
\label{sec:Data_Augmentation}

To prevent overfitting and improve general system robustness and generalization ability, we utilize several data augmentation techniques. All augmentations are applied on-the-fly during training. We can split the data augmentation methods into augmentation techniques that are performed at either the waveform and spectrogram level. At the waveform level, we choose to only utilize the audio channel swapping (ACS) method as proposed by Wang et al. in \cite{wang2023four}. The ACS augmentation method swaps the audio channels around, thus creating new DOA representations. The ACS method also strongly improves SELD in real recordings as noted in \cite{niu2023experimental} as they maintain the reverberation and diffuse effects of real life recordings. As such, we only apply the ACS augmentation method to data from the STARSS23 dataset.

At the spectrogram level, we use SpecAugment \cite{Park_2019_SpecAug}, Mixup \cite{zhang2018mixup} along both the time and frequency domain, as well as frequency shifting \cite{Nguyen2022_SALSALite}. For spectrogram level augmentations, we adopt a similar augmentation policy as the one detailed by Wang et al. in \cite{wang2022cyclicaugment}. In their work, the authors proposed a data augmentation approach to dynamically configure the magnitude of augmentation policies with a cosine annealing scheduler. For our training setup, we also reduce the magnitude of spectrogram level augmentations as the learning rate decays. However, we still apply the ACS augmentation method at the same probability throughout training. 

\subsection{Network Architecture}

\begin{figure}[t]
    \centering
    \includegraphics[width=0.85\columnwidth]{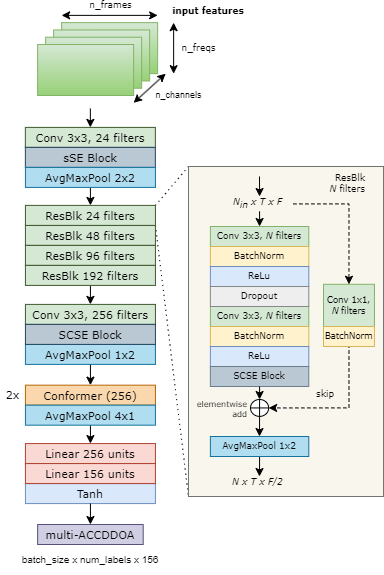}
    \caption{Proposed system architecture with added sSE and SCSE blocks}
    \label{fig:proposed-model}
\end{figure}

We use the ResNet-Conformer system, which was used by most top-ranking teams in the previous year's DCASE Challenge Task 3. The conformer architecture was first proposed by Gulati et al. and refers to convolution-augmented transformers \cite{gulati2020conformer}. A conformer block comprises four modules stacked consecutively: a feed-forward module, a self-attention module, a convolution module, and a second feed-forward module. Overall, the attention module helps to recognize global features, while the convolution module helps to identify local features. Combining the two yields SOTA results in the field of speech recognition, and has been used in SELD works as well to positive effect, as seen from the results of DCASE Challenge 2023 Task 3.

To further improve performance, we utilized a deeper, modified ResNet-18 structure for the encoding layer preceding the conformer blocks as detailed by Niu et al. in \cite{niu2023experimental}. In the proposed system, we apply pooling operations only along the frequency dimension after each ResNet block. The authors also noted that the conformer architecture benefits from a higher temporal resolution, and find that the information of impulse or impact sound events may be affected by early time pooling operations. As such, we choose to apply the time pooling operation after the conformer blocks to preserve temporal information within the network. We also apply the \textit{tanh} activation function at the end of our model, which is only possible due to the distance values being scaled to be within $[-1, 1]$. Our system utilizes the multi-ACCDDOA output format as proposed in \cite{krause2024sound_dcasebaseline}. We train the model using the mean squared error (MSE) loss function with auxiliary duplicating permutation invariant training, as described by Shimada et al. \cite{shimada_multiaccdoa}. The full proposed model architecture is shown in Figure \ref{fig:proposed-model}. The pooling operations are done along the \textit{time} $\times$ \textit{frequency} dimensions. We utilize AvgMaxPool, which refers to the arithmetic sum of the outputs of an average pooling and max pooling operation on an input feature map. We adopt this method from \cite{kong2020panns} to maximize the benefit of both forms of pooling operations. 

\subsection{Squeeze-and-Excitation Blocks}

\begin{figure}[t]
    \centering
    \includegraphics[width=0.99\columnwidth]{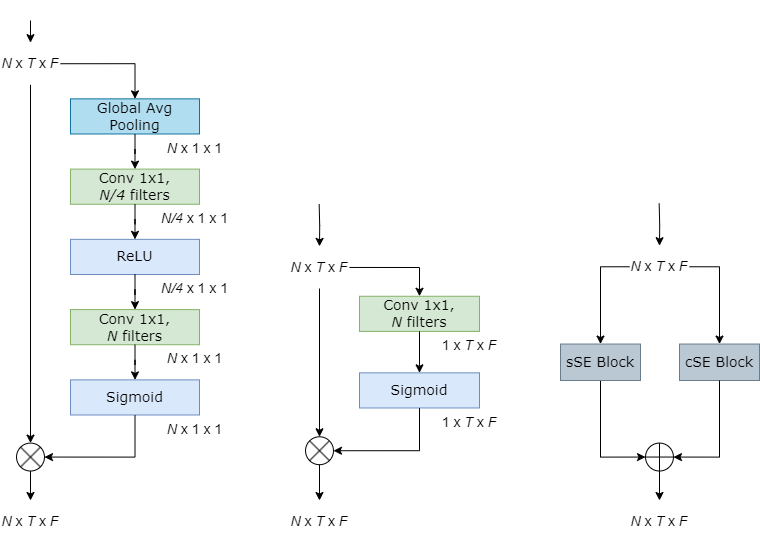}
    \caption{Block diagram of the cSE block (left), sSE block (middle) and SCSE block (right)}
    \label{fig:all_SCSEBlocks}
\end{figure}

One of the key reasons behind the effectiveness of the SALSA suite of features is the channel-wise alignment of the time frequency (TF) bins of both the spectral and spatial features. In SALSA, a linear frequency scale is used instead of the typical log-mel frequency scale, which helps preserve frequency information without any form of aggregation. After linearly compressing the frequency bins above 9kHz in SALSA, we end up with 200 frequency bins, as compared to the 64 or 128 Mel bins used in typical Mel-spectrogram features. In order to make full use of the expansive frequency information, we implement additional functional layers to improve the system's ability to learn inter-channel spatial information. 

Squeeze-and-Excite (SE) blocks were first introduced by Hu et al. as additional layers to CNNs \cite{hu2018squeeze} for image segmentation work. SE blocks work by first aggregating global spatial information into a channel descriptor. This is usually done through global average pooling to form a vector with a length equal to the number of convolutional channels. After which, a self-gating mechanism is used to ``excite" the vector, which is then used to rescale the output feature maps. This can be seen as a form of ``channel attention", where the SE block aims to actively scale the channel importance and model the inter-channel dependencies. In later work, Roy et al. introduced the idea of channel- and spatial-wise SE blocks \cite{roy2018concurrent}. The original SE block proposed by Hu et al. can be seen as a spatial squeeze and channel excitation block (cSE). Roy et al. instead proposed its counterpart, which is the channel squeezing and spatial excitation block (sSE). For the sSE block, the goal is to learn the important spatial mapping of the input feature maps and can be seen as a form of ``spatial attention". For the case of SELD, we propose that the sSE block can be seen as a method to help the system focus on the important TF bins across all channels. The authors also propose the concurrent spatial and channel SE block (SCSE) which is a simple element-wise addition of the sSE and SE excitations. In the output of the SCSE block, a location will receive a higher activation if it gets a higher importance from both the channel and spatial recalibration, or excitation. The SCSE block aims to encourage the network to learn more meaningful feature maps that are relevant both spatially and channel-wise. We show the workings of the cSE, sSE and SCSE block in Figure \ref{fig:all_SCSEBlocks}.

In our proposed system, we add sSE blocks into the stem blocks of the ResNet encoding layer, as this is the section of the system where all information across the time and frequency dimensions are preserved without aggregation. Our motive is for the system to learn the important inter-channel information maps at the highest possible frequency resolution. After which, we include SCSE blocks into the subsequent residual layers of the ResNet to learn meaningful spatial and channel representations. We follow the work of \cite{naranjoalcazar2021SELDSE} and squeeze the number of channels by a factor of 4 in each cSE block in our systems.

\subsection{Training}

For training, we first segment the input audio signal into 5 second segments without overlapping before converting them into SALSA features. We train the model for 200 epochs using the Adam optimizer, with a peak learning rate of $5\times10^{-4}$ and a batch size of 32, using the transformer learning rate scheduler that was first proposed in \cite{vaswani2023attention}. We follow the implementation of the baseline system and evaluate the model on the test split of the development set of the STARSS23 dataset, saving the model with the best validation location-dependent F1-Score as our final model. Furthermore, we further fine-tune the models for another 50 epochs on only real audio recordings after the initial training. 


\section{RESULTS}

We first compare the performance improvements of each of our optimization methods on the baseline system using the training split of the STARSS23 dataset. However, we did not compare the usage of the SALSA features using the baseline system, as the baseline system is optimized for the 64 Mel bins of the Mel-spectrogram input features. We instead only compare the usage of distance scaling (DS), spectrogram level augmentations (Aug) and the audio channel swapping augmentation method (ACS). We evaluate each model on the test split of the development set of the STARSS23 dataset, using the DCASE Challenge 2024 macro-averaged evaluation metrics that take SDE into account\footnote{https://dcase.community/challenge2024/task-audio-and-audiovisual-sound-event-localization-and-detection-with-source-distance-estimation\#evaluation}. The jackknife estimation method was used when calculating the metrics during evaluation. The full breakdown of our experiments is presented in Table \ref{tab:Baseline_Comparison}, with the first row of the table denoting the results of the baseline system without any modifications. We also introduce an aggregated SELDDE error function that is similar to the SELD aggregated error function used in previous years' DCASE Challenge Task 3 \cite{politis2020overview}. The SELDDE error, $\mathcal{E}_{\text{\tiny{SELDDE}}}$, is given by,

\begin{equation}
    \mathcal{E}_{\text{\tiny{SELDDE}}} = \frac{1}{3}((1-{\text{F}}_{20^\circ}) + \frac{\text{LE}_{\text{CD}}}{180^\circ} + \text{RDE}_{\text{CD}}),
\end{equation}

where $\text{F}_{20^\circ}$, $\text{LE}_{\text{CD}}$ and $\text{RDE}_{\text{CD}}$ represent the location-dependent F1-score, class-dependent localization error and class-dependent relative distance error respectively. This SELDDE error function is not used for any official DCASE evaluation, but can be seen as a measure of the system's overall SELDDE performance.

\begin{table}[t]
\caption{SELDDE performance of the baseline model using various optimization methods}
\centering
    \begin{tabularx}{\columnwidth}{|>{\centering\arraybackslash}c >{\centering\arraybackslash}c >{\centering\arraybackslash}c  >{\centering\arraybackslash}X >{\centering\arraybackslash}X >{\centering\arraybackslash}X >{\centering\arraybackslash}c|}
    \hline
        ACS & Aug & DS & $\text{F}_{20^\circ} \uparrow$ & $\text{LE}_{\text{CD}} \downarrow$ & $\text{RDE}_{\text{CD}}$$\downarrow$ & $\mathcal{E}_{\text{\tiny{SELDDE}}} \downarrow$ \\\hline
        & & & 13.1\% & $36.9^\circ$ & 33.0\% & 0.468 \\
        & & \Checkmark & 13.6\% & $33.2^\circ$ & 31.0\% & 0.453\\
        & \Checkmark & & 15.0\% & $29.4^\circ$ & 27.3\% & 0.429\\
        & \Checkmark & \Checkmark & 15.9\% & $25.5^\circ$ & 31.6\% & 0.433 \\
        \Checkmark & \Checkmark & \Checkmark & 16.8\% & $21.1^\circ$ & 31.0\% & 0.420\\\hline
    \end{tabularx}
    \label{tab:Baseline_Comparison}
\end{table}

We show that the simple distance scaling trick can help the system to train with a more stable loss function, which results in slightly improved SED, DOA and SDE metrics. The usage of various data augmentation techniques also help to improve the system's ability to generalize, resulting in noticeable improvements in all SELDDE metrics. Combining the two methods also improves SED, DOA, and SDE performance over the baseline by a substantial margin. As reported in \cite{wang2023four, niu2023experimental}, the ACS method does help to generate additional realistic DOA representations, which helps to improve DOA metrics significantly. 

\begin{table}[t]
\centering
\caption{SELDDE performance of each of our submitted systems in comparison with the baseline system (BL) using the FOA format}
    \begin{tabularx}{\columnwidth}{|>{\centering\arraybackslash}X >{\centering\arraybackslash}X >{\centering\arraybackslash}X >{\centering\arraybackslash}X  >{\centering\arraybackslash}X|}
    \hline
        System & $\text{F}_{20^\circ}\uparrow$ & $\text{LE}_{\text{CD}} \downarrow$ & $\text{RDE}_{\text{CD}} \downarrow$ & $\mathcal{E}_{\text{\tiny{SELDDE}}} \downarrow$ \\\hline
        BL & 13.1\% & $36.9^\circ$ & 33.0\% & 0.468\\
        BL$+$ & 19.0\% & $29.1^\circ$ & 30.6\% & 0.426\\\hline
        A & \textbf{33.9\%} & $\textbf{20.4}^\circ$ & 30.4\% & \textbf{0.359} \\
        B & 33.8\% & $21.4^\circ$ & \textbf{30.0\%} & 0.360 \\
        C & 32.7\% & $22.9^\circ$ & 30.1\% & 0.367 \\
        D & 32.7\% & $20.6^\circ$ & 31.1\% & 0.366 \\\hline
        
    \end{tabularx}
    \label{tab:Full_Results}
\end{table}

Next, we describe the performance of each of our systems submitted in the test split of the development set of the STARSS23 dataset in Table \ref{tab:Full_Results}. BL stands for the performance of the baseline system, while BL$+$ represents the baseline system trained on the dataset used for submission. All model submissions use the underlying ResNet-Conformer architecture. System A uses all SE blocks as shown in Figure \ref{fig:proposed-model}. System B removes the final SCSE block from the tail of the system. System C removes all SCSE blocks from the ResNet blocks, leaving only the sSE and SCSE blocks in the stem and tail of the system, respectively. Lastly, system D removes all sSE and SCSE blocks from the system entirely, resulting in the ``bare-bones" ResNet-Conformer system.

Overall, the inclusion of SE blocks into the system architecture improves the SELDDE performance, as demonstrated by the improved metrics of systems A and B. We conjecture that it could due to the SE blocks introducing slight forms of spatial attention along the TF bins to better identify classes which results in the improved SED F1-scores. The slight difference in the system performances of systems A and B does also seem to suggest that the placement of each SE block is crucial to the best performance of the system. We leave the study for the optimal placement and positioning of SE blocks for future study. 




\section{Conclusion}
\label{sec:Conclusion}

This technical report presents our proposed system architecture of combining ResNets, Conformer modules and Squeeze-and-Excitation blocks for SELDDE tasks. We also reintroduce SALSA features for polyphonic SELD tasks and forms of channel- and spatial-wise attention modules within the convolutional layers. Similarly, for future work, we hope to present another suite of hand-crafted features dedicated to SELDDE performance. In general, our proposed approach brings about significant improvements over the baseline method. We recognize that we are unable to provide a full, detailed evaluation of all our proposed improvements in this report. We leave this further ablation study for future work. 

\section{ACKNOWLEDGEMENTS}
\label{sec:acknowledgements}

This research is supported by the Ministry of Education, Singapore under its Academic Research Fund Tier 2 (MOE-T2EP20221-0014).

\bibliographystyle{IEEEtran}
\bibliography{refs}

\end{sloppy}
\end{document}